# Superconductivity in $CuIr_{2-x}Al_xTe_4$ telluride chalcogenides[*]

Dong Yan (严冬)[1,4 #], Lingyong Zeng (曾令勇)[1 #], Yijie Zeng (曾宜杰)[2,5], Yishi Lin (林一石)[3], Junjie Yin (殷俊杰)[2], Meng Wang (王猛)[2], Yihua Wang (王熠华)[3], Dao-Xin Yao (姚道新)[2] and Huixia Luo (罗惠霞)[1†]

[1] *School of Materials Science and Engineering, State Key Laboratory of Optoelectronic Materials and Technologies, Key Lab of Polymer Composite & Functional Materials, Guangzhou Key Laboratory of Flexible Electronic Materials and Wearable Devices, Sun Yat-Sen University, No. 135, Xingang Xi Road, Guangzhou, 510275, P. R. China*

[2] *Center for Neuron Science and Technology, School of Physics, Sun Yat-Sen University, No. 135, Xingang Xi Road, Guangzhou, 510275, P. R. China*

[3] *State Key Laboratory of Surface Physics and Department of Physics, Fudan University, Shanghai 200433, P. R. China; Shanghai Research Center for Quantum Sciences, Shanghai 201315, China*

[4] *Key Laboratory of Functional Molecular Solids, Ministry of Education, College of Chemistry and Materials Science, Anhui Normal University, Wuhu, 241002, China*

[5] *College of Science, Hangzhou Dianzi University, Hangzhou 310018, P. R. China*

[#] These authors contributed equally to this work.

The relationship between charge-density-wave (CDW) and superconductivity (SC), two vital physical phases in condensed matter physics, has always been the focus of scientists' research over the past decades. Motivated by this research hotspot, we systematically studied the physical properties of the layered telluride chalcogenide superconductors $CuIr_{2-x}Al_xTe_4$ ($0 \leq x \leq 0.2$). Through the resistance and magnetization measurements, we found that the CDW order was destroyed by a small amount of Al doping. Meanwhile, the superconducting transition temperature ($T_c$) kept changing with the change of doping amount and rose towards the maximum value of 2.75 K when $x = 0.075$. The value of normalized specific heat jump ($\Delta C/\gamma T_c$) for the highest $T_c$ sample $CuIr_{1.925}Al_{0.075}Te_4$ was 1.53, which was larger than the BCS value of 1.43 and showed that bulk superconducting nature. In order to clearly show the relationship between SC and CDW states, we propose a phase diagram of $T_c$ *vs.* doping content.

[*] H. X. Luo acknowledges the financial support by the National Natural Science Foundation of China (Grants No.11922415), Guangdong Basic and Applied Basic Research Foundation (2019A1515011718), and the Pearl River Scholarship Program of Guangdong Province Universities and Colleges (20191001). Y. Zeng and D. X. Yao are supported by the National Natural Science Foundation of China (Grants No. 11974432), NKRDPC-2018YFA0306001, NKRDPC-2017YFA0206203. D. Yan acknowledges the financial support by National Key Laboratory Development Fund (No. 20190030). Y. H. Wang would like to acknowledge partial support by the Ministry of Science and Technology of China under Grant No. 2017YFA0303000, NSFC Grant No. 11827805 and Shanghai Municipal Science and Technology Major Project Grant No. 2019SHZDZX01. M. Wang was supported by the National Natural Science Foundation of China (Grants No. 11904414, 12174454) and National Key Research and Development Program of China (Grants No. 2019YFA0705702).

[†] Corresponding author. E-mail: *luohx7@mail.sysu.edu.cn*

## 1. Introduction

The continuous suppression of metal-insulator (MI), charge-density-wave (CDW) transition and magnetism, etc. leading to the occurrence of superconductivity (SC) in the proximity of such quantum states has garnered great interest and widespread study in solid-state physics. [1-4] The phase diagrams of unconventional high-temperature (high-$T_c$) cuprates and iron-based superconductors exemplified such phenomena. [5-11] Nevertheless, how SC emerges in these high-$T_c$ superconductors are intricate and remains extremely puzzling. It is still essential to sort out the interplay between the SC and the other quantum states, promoting further understanding of the mechanism of high-$T_c$ superconductors.

Low-dimensional layered transition metal dichalcogenides (TMDs) are other opportune material platforms for the exploration of various quantum instabilities, [12-21] especially the interplay between SC and CDW, in which the CDW order refers to condensate with periodic modulations of the crystalline lattice and conduction electron density in real space. Typically, the SC can be induced and a dome-shape superconducting phase diagram is formed upon suppressing the CDW order, [22-27] which is highly similar to the phase diagrams of unconventional high-$T_c$ cuprates and iron-based superconductors. Despite overall phase diagram similarities, there are significant differences especially between their mechanisms, where unconventional high-$T_c$ Fe-based and cuprate superconductors cannot be well explained by Bardeen-Cooper-Schrieffer (BCS) theory but most of the TMD superconductors can be explained by BCS theory. It has been generally considered that the collapse of CDW state accompanied by the improvement of superconducting transition temperature ($T_c$) is in reference to the abrupt enhancement of density of states (DOS) around the Fermi level $N(E_F)$ in the conventional superconductors owning to CDW state gaps out some regions of the Fermi surface. [1,2] The formation of $Cu_xTiSe_2$ ($0 \leq x \leq 0.1$) from intercalating the Cu into 1T-TiSe$_2$ exemplified a vivid phase diagram in the TMDs family, [2] and further evoked the continuous interest in searching for new superconductors in TMD materials by gating, adding physical pressures, chemical doping or point contact method. [22-38] For example, Cu-intercalation 2H-TaS$_2$ forms $Cu_xTaS_2$ ($0 \leq x \leq 0.12$) and displays an enhancement of $T_c$ from 0.8 K ($x = 0$) to 4.5 K ($x = 0.04$). [34] In addition, experiments show that the $T_c$ of WTe$_2$ can be increased to 7 K under an applied physical pressure of 16.8 GPa. [35] Moreover, manifold phase transitions from MI to metal show in 1T-TaS$_2$ thin flakes with collapses of CDWs, and finally SC is induced by ionic gating. [36]

Currently, CuIr$_2$Te$_4$, adopting a NiAs defected structure of trigonal symmetry with the space group $P$-3$m$1, has been found to exhibit the coexistence of the CDW-like transition ($T_{CDW}$ = 250 K on heating and 186 K on cooling) and SC ($T_c$ = 2.5 K). Besides, recent electronic structure calculations have unveiled that the Fermi energy is mostly derived from Ir 5$d$ and Te 5$p$ orbitals. [39] More recently, experiments have documented that both CDW and SC performances of CuIr$_2$Te$_4$ can be modified via 3$d$, 4$d$ transition metals (e.g., Ti and Ru) and 4$p$, 5$p$ dopants (e.g., Se and I). Dome-shape phase diagrams with respect to $T_c$ vs. doping amount associated with the suppression of CDW have been found in the CuIr$_{2-x}$Ru$_x$Te$_4$ and CuIr$_{2-x}$Ti$_x$Te$_4$ systems, but dome-shape diagrams crowded in the middle doping region with two CDW regions at two sides in Se and I-doped systems. [40-43] Furthermore, the non-magnetic element Al is usually selected as the dopant because $Al^{3+}$ ion has a closed shell electron configuration and a clear oxidation state. The transport properties of Al-doped high-$T_c$ copper-based superconductors (e.g. $SmBa_2Cu_{3-x}Al_xO_{6+\delta}$, $YBa_2Cu_{3-x}Al_xO_7$) has been widely studied. [44, 45] And nano Al has been used to improve the critical current in $MgB_2$ superconductor. [46] Therefore, it will be interesting to explore the effect of 3$s$ dopants (e.g., Al) on the CDW and SC in $CuIr_2Te_4$.

In this work, we prepared the polycrystalline $CuIr_{2-x}Al_xTe_4$ ($0 \leq x \leq 0.2$) compounds successfully by a solid-state reaction method. Our results demonstrate that the CDW order can be completely suppressed within a fine-tuned Al-doped content as a result of the improvement of $T_c$. $T_c$ initially increases with the rise of doping amount until $x = 0.075$ and reaches highest value of 2.75 K, eventually forming a dome-phase like electronic phase diagram. The acquisition of the $CuIr_{2-x}Al_xTe_4$ ($0 \leq x \leq 0.2$) system also provides some enlightenment for the search of new superconductors.

2. **Experimental Methods**

**Synthesis:** Polycrystalline specimens of $CuIr_{2-x}Al_xTe_4$ ($0 \leq x \leq 0.2$) were prepared through the classical solid-state phase reaction. First, the Cu (99%, Alfa Aesar), Ir (99.9%, Macklin), and Al (99.95%, Aladdin) powder, and Te lump (99.999%, Alfa Aesar) with an element ratio of Cu : Ir : Al : Te = 1 : 2-$x$ : $x$ : 4.05 was sealed in quartz tubes, then put them in a muffle furnace with the ramping rate 1 ºC/min to 850 ºC and maintain the temperature for 5 days. The resulting samples were annealed at 850 ºC for 4 days with a heating rate 1 ºC/min.

**Instruments:** Powder X-ray diffraction (PXRD) MiniFlex, Rigaku with Cu $K\alpha$1 radiation was used to examine the crystal structure and phase purity for $CuIr_{2-x}Al_xTe_4$ ($0 \leq x \leq 0.2$) compounds. FULLPROF software suite was used to determine the cell parameters based on Thompson-Cox-Hastings pseudo-Voigt peak shapes model. Measurements of the temperature-dependence of electrical resistivity, specific heat, and magnetic susceptibilities (M(T, H)) were performed by a DynaCool Quantum Design Physical Property Measurement System (PPMS, Quantum Design, Inc.)

3. **Results and Discussion**

The PXRD patterns of $CuIr_{2-x}Al_xTe_4$ compounds are presented in **Fig. 1a**. PXRD analysis shows that Al concentration is limited up to 0.2 since $Al_2Te_3$ impurity is found with further increasing Al content. As Al doping content increases, the (001) peak moves to the right, which can be verified by the decrease of lattice constants $a$ and $c$ with increasing $x$ (**Fig. 1c**), reflecting the compression of the $CuIr_2Te_4$ unit cell. As illustrated in **Fig. 1c**, clearly, both lattice constants ($a$ and $c$) and $c/a$ singly reduce with increasing Al concentration. It is found that $a$ and $c$ reduce from 3.9397(5) and 5.3965(3) Å for the pristine sample to 3.9264(1) and 5.3757(2) Å ($x = 0.2$) in $CuIr_{2-x}Al_xTe_4$, respectively. The detailed refinement of the selected sample $CuIr_{1.925}Al_{0.075}Te_4$ is displayed in **Fig. 2a**. Most of the diffraction peaks have been indexed in terms of trigonal symmetry with a space group $P$-3$m$1 (No. 164) and some small peaks indexed for tiny unreacted Ir are also detected. The illustration shows that the disordered trigonal structure, in which the Cu is inserted between two-dimensional (2D) IrTe$_2$ layers, Ir partial substituting by Al simultaneously (see **Fig. 2b and 2c**).

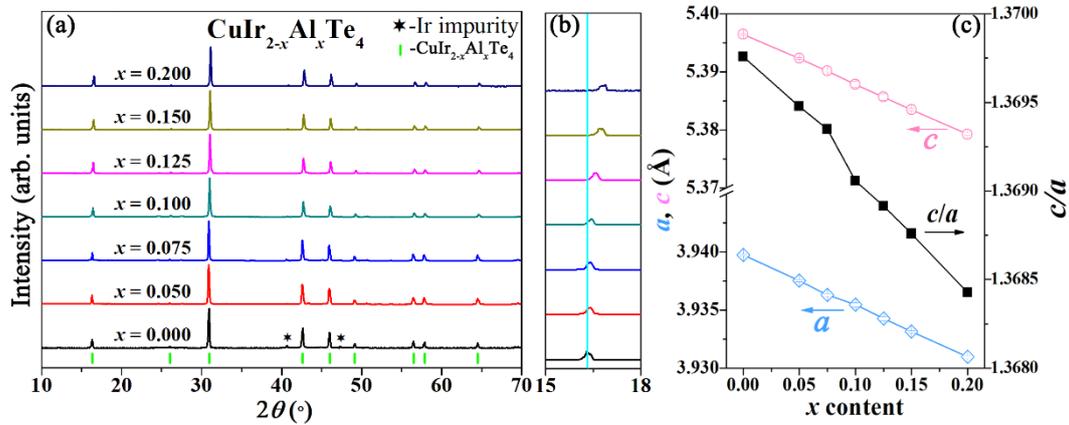

**Fig. 1** (a) and (b) displays PXRD patterns of the CuIr$_{2-x}$Al$_x$Te$_4$ ($0 \leq x \leq 0.2$) compounds. (c) The evolution of lattice constants for CuIr$_{2-x}$Al$_x$Te$_4$ ($0 \leq x \leq 0.2$).

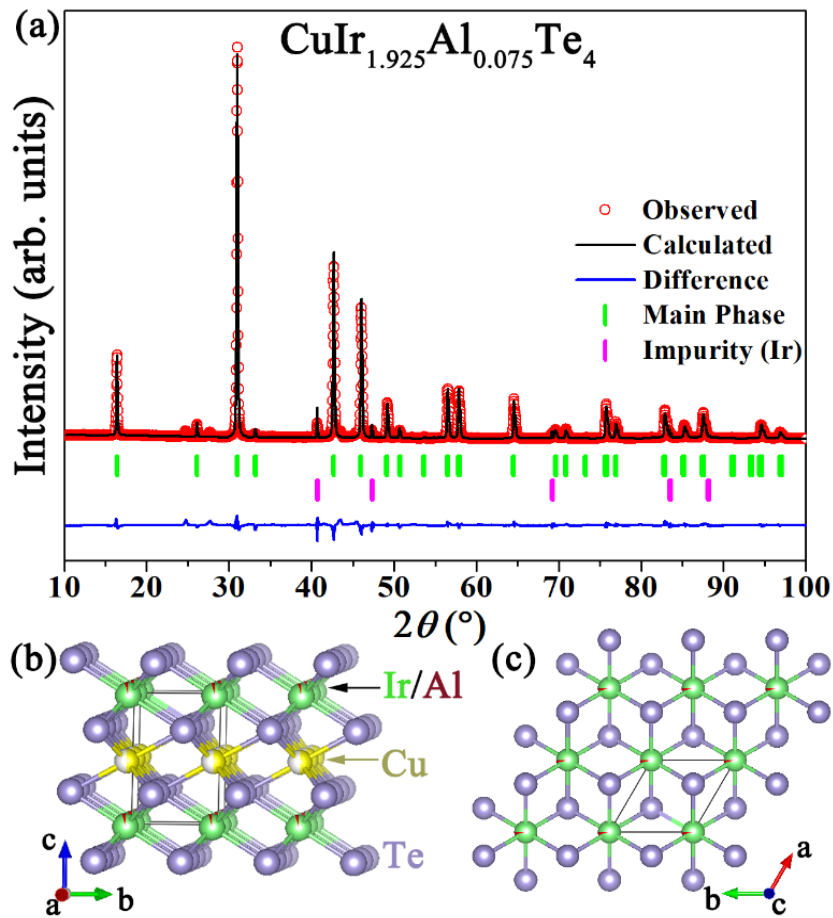

**Fig. 2** (a) Refinements of CuIr$_{1.925}$Al$_{0.075}$Te$_4$ polycrystalline sample. (b) and (c) show the crystal structure of CuIr$_{2-x}$Al$_x$Te$_4$ in different direction views.

**Table 1.** $T_c$ and RRR of CuIr$_{2-x}$Al$_x$Te$_4$ ($0 \leq x \leq 0.2$) compounds. The $T_c$ is determined by using

50% normal state resistance criterion.

| x content | $T_c$ (K) | RRR |
|---|---|---|
| 0 | 2.50 | 4.149 |
| 0.05 | 2.74 | 4.367 |
| 0.075 | 2.75 | 4.444 |
| 0.1 | 2.72 | 4.115 |
| 0.125 | 2.71 | 2.865 |
| 0.15 | 2.41 | 3.846 |
| 0.2 |  | 2.342 |

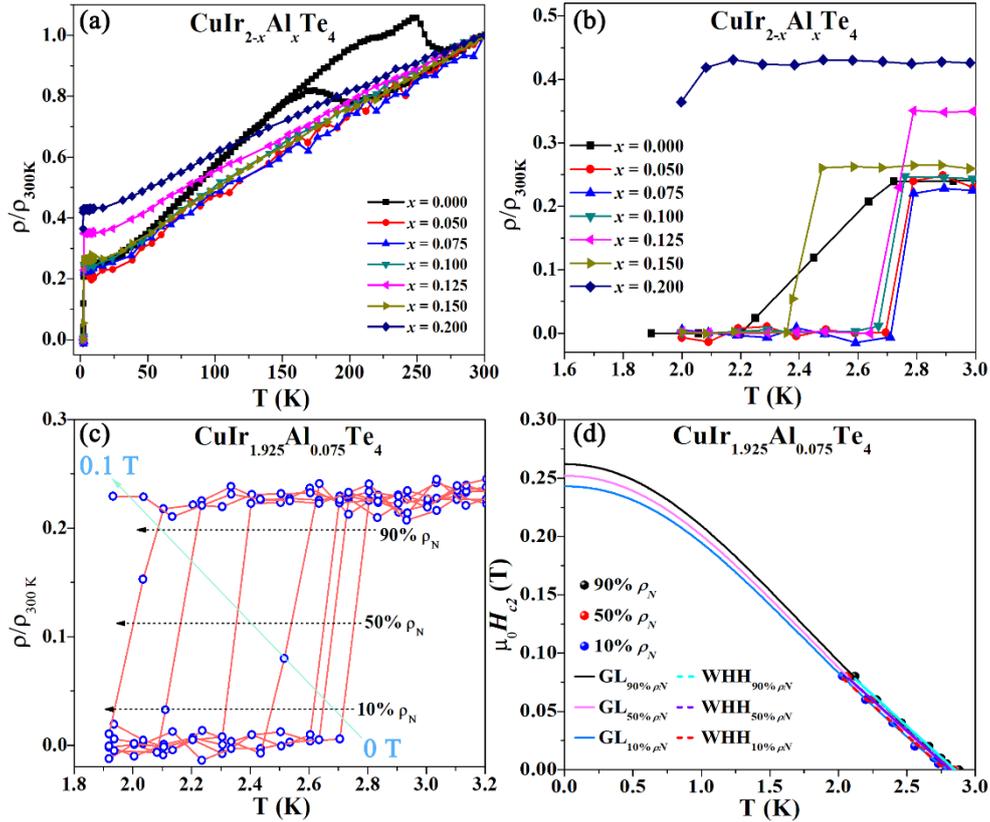

**Fig. 3** (a) Temperature-dependence of resistivity for the polycrystalline $CuIr_{2-x}Al_xTe_4$ ($0 \leq x \leq 0.2$) compounds. (b) The superconducting transition at 1.6 - 3.0 K for the polycrystalline $CuIr_{2-x}Al_xTe_4$ ($0 \leq x \leq 0.2$) compounds. (c) Temperature dependences of resistivity for the polycrystalline $CuIr_{1.925}Al_{0.075}Te_4$ sample under various applied magnetic fields. (d) Temperature dependence of $H_{c2}$ for the polycrystalline $CuIr_{1.925}Al_{0.075}Te_4$ sample.

The temperature-dependent normalized resistivity ($\rho/\rho_{300K}$) of the $CuIr_{2-x}Al_xTe_4$ ($0 \leq x \leq 0.2$) compounds is exhibited in **Fig. 3a**. Besides, our previous finding indicates that there is no structure transition for the pristine sample. Thus, we propose that the normalized resistivity with cooling and heating exhibited a distinct hysteresis associated with the formation of CDW-like transition for the pristine $CuIr_2Te_4$ sample. It is worth mentioning that no signature of the abnormal hump can be observed above $T_c$ in $\rho/\rho_{300K}$ of the Al doping samples $CuIr_{2-x}Al_xTe_4$ ($0 < x \leq 0.2$), indicating the CDW-like transition can be suppressed with subtle Al substitution for Ir, companying with the increment of $T_c$. The resistivity data for Al-doped samples show a metallic behavior.

We can observe sharp drops of $\rho$(T) for the $CuIr_{2-x}Al_xTe_4$ ($0 \leq x \leq 0.15$) below 3.0 K (see **Fig. 3b**), which represent the outset of the superconducting state. The transition width of Al-doped compounds is much narrower than that of the pristine $CuIr_2Te_4$. The $T_c$ and residual resistivity ratio (RRR = R(300K)/R(3K)) of $CuIr_{2-x}Al_xTe_4$ ($0 \leq x \leq 0.2$) samples summarized in **Table 1**. As Al content increases, both $T_c$ and RRR increase to the highest value at $x = 0.075$, then decrease for further Al doping. For $0.075 \leq x \leq 0.5$, the RRR value sharply decreases form 4.44 for $x = 0.075$ to 2.342 for $x = 2.34$. This phenomenon indicates that high Al content induces significant disorder in the polycrystalline $CuIr_{2-x}Al_xTe_4$ series.

**Table 2.** Comparison of physical properties of $CuIr_2Te_4$-based superconductors.

| Compound | $CuIr_{1.925}Al_{0.075}Te_4$ | $CuIr_{1.925}Ti_{0.075}Te_4$ [33] | $CuIr_{1.95}Ru_{0.05}Te_4$ [32] | $CuIr_2Te_4$ [31] |
|---|---|---|---|---|
| $T_c$ (K) | 2.75 | 2.84 | 2.79 | 2.50 |
| $\gamma$ (mJ mol$^{-1}$ K$^{-2}$) | 12.12 | 14.13 | 12.26 | 12.05 |
| $\beta$ (mJ mol$^{-1}$ K$^{-4}$) | 2.20 | 2.72 | 1.87 | 1.97 |
| $\Theta_D$ (K) | 183.5(1) | 170.9(1) | 193.6(2) | 190.3(1) |
| $\Delta C/\gamma T_c$ | 1.53 | 1.34 | 1.51 | 1.50 |
| $\lambda_{ep}$ | 0.66 | 0.64 | 0.65 | 0.63 |
| $N(E_F)$ (states/eV f.u) | 3.13 | 3.67 | 3.15 | 3.10 |
| $\mu_0 H_{c2}$(T) (50 % $\rho_N$ WHH theory) | 0.191 | 0.212 | 0.247 | 0.12 |
| $\mu_0 H^P$(T) | 5.12 | 5.28 | 5.24 | 4.65 |
| $\mu_0 H_{c1}$(T) | 0.060 | 0.095 | 0.098 | 0.028 |
| $\xi_{GL}(0)$ (nm) (50 % $\rho_N$ WHH theory) | 40.4 | 39.3 | 36.3 | 52.8 |

To evaluate the superconducting properties of $CuIr_{1.925}Al_{0.075}Te_4$ compound with the highest $T_c$ in detail, we have measured the resistivity under various applied magnetic fields. $T_c$s shift to lower temperature with the increasing applied field (see in **Fig. 3c**). The upper critical field $\mu_0 H_{c2}(0)$ is obtained from Werthamer-Helfand-Hohenberg (WWH) and Ginzburg-Landau (GL) models. The upper critical field $\mu_0 H_{c2}$(T) values of $CuIr_{1.925}Al_{0.075}Te_4$ at zero temperature can be extrapolated by GL formula using the data deriving from 10, 50, and 90 % criteria of $\rho_N$ (normal-state resistivity). The GL formula is shown as follows: $\mu_0 H_{c2} = \mu_0 H_{c2}(0) * \frac{1-(T/T_c)^2}{1+(T/T_c)^2}$. [47] The $\mu_0 H_{c2}(0)$ values of 0.235, 0.250, and 0.255 T for $CuIr_{1.925}Al_{0.075}Te_4$ can be determined using the data 10, 50 and 90 % criteria of $\rho_N$, respectively. Furthermore, we can acquire $\mu_0 H_{c2}(0)$ values of 0.180, 0.191, and 0.193 T for $CuIr_{1.925}Al_{0.075}Te_4$ corresponded to the data 10, 50, and 90 % criteria of $\rho_N$, respectively, via using the WHH formula $\mu_0 H_{c2} = -0.693 T_c \frac{dH_{c2}}{dT_c}$ for the dirty limit SC. [48] The slopes, $dH_{c2}/dT$, are obtained from linear fitted $CuIr_{1.925}Al_{0.075}Te_4$ sample for data 10, 50 and 90 % criteria of $\rho_N$. Additionally, we fit the Pauli limiting field ($\mu_0 H^P$(T)) by the formula $\mu_0 H^P$ (T) = $1.86 T_c$, which is shown in **Table 2**. Then, the Ginzburg-Landau coherence length $\xi_{GL}(0)$ can be derived from this equation $\mu_0 H_{c2}(T) = \frac{\phi_0}{2\pi\xi_{GL}^2}$, where $\phi_0$ represents the flux quantum. For example, $\xi_{GL}(0)$ of CuIr$_{1.925}$Al$_{0.075}$Te$_4$ is calculated to be 40.4 nm via using 50 % criteria of $\rho_N$ data based on WHH model. **Table 2** summarizes all the relative experimentally measured and estimated parameters.

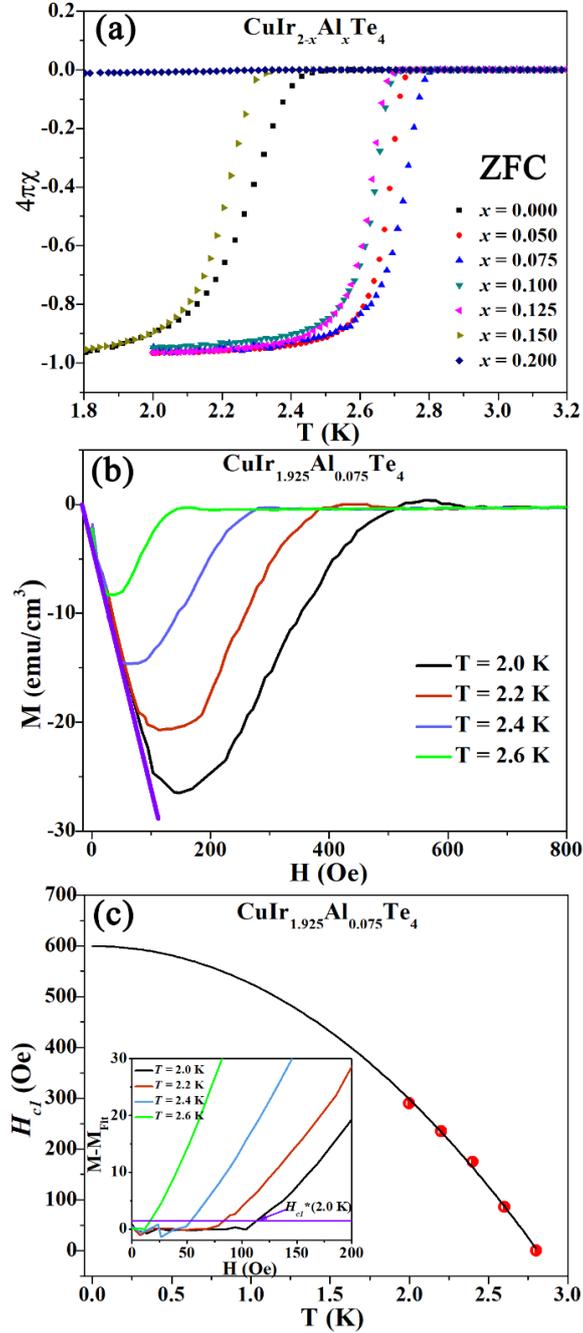

**Fig. 4** (a) Magnetic susceptibility for CuIr$_{2-x}$Al$_x$Te$_4$ (0 ≤ x ≤ 0.2) compounds under 30 Oe magnetic field with zero-field-cooling (ZFC) mode. (b) The magnetization M(H) in a temperature range between 2 - 2.6 K of CuIr$_{1.925}$Al$_{0.075}$Te$_4$. (c) Temperature dependence of $H_{c1}$. Inset: Difference between M and M$_{Fit}$ in a temperature range between 2 - 2.6 K of CuIr$_{1.925}$Al$_{0.075}$Te$_4$.

The superconducting transition has also been confirmed by the magnetic susceptibility data (**Fig. 4a**) with strong diamagnetic signals under zero-field-cooling (ZFC) mode. The superconducting volume fraction of the $CuIr_{2-x}Al_xTe_4$ ($0 \leq x \leq 0.15$) compounds can be calculated around 95 %, which manifests high purity of the polycrystalline $CuIr_{2-x}Al_xTe_4$ samples. Next, we examine the lower critical fields ($\mu_0H_{c1}$) by the field-dependent magnetic susceptibility M(H) measurements in a temperature range between 2 - 2.6 K. **Fig. 4b** presents the magnetization (M-H) curves at different temperatures below $T_c$ of the representative $CuIr_{1.925}Al_{0.075}Te_4$. As shown in the purple line in **Fig. 4b**, the M(H) isotherms show a linear relationship with the magnetic field (H) at low magnetic fields, indicating it is a type-II superconductor. We can extract the demagnetization factor ($N$) values following the expression: $N = 4\pi\chi_V + 1$, where $\chi_V = \frac{dM}{dH}$ represents the linearly fitted slope. The calculated $N$ value for $CuIr_{1.925}Al_{0.075}Te_4$ is about 0.53. We can use the expression $M_{fit} = m + nH$ to fit the experimental data at low magnetic fields, where $m$ represents the intercept and $n$ stands for the slope of linear fitting from the low magnetic field M(H) data. The inset of **Fig. 4c** displays the (M - $M_{fit}$) data *vs.* H. Then, we can fit the $\mu_0H_{c1}$(T) by the expression $\mu_0H_{c1}(T) = \frac{\mu_0H_{c1}^*(T)}{1-N}$, where $\mu_0H_{c1}^*$ is the intersection point between the M - $M_{fit}$ vs. H curves and the field (purple line in the inset of **Fig. 4c**), which deviates by ∼ 1 % above the fitted data ($M_{fit}$) as customary. The temperature dependence of $\mu_0H_{c1}$(T) for $CuIr_{1.925}Al_{0.075}Te_4$ is displayed in the main panel of **Fig. 4c**. Consequently, we can further acquire the $\mu_0H_{c1}$(T) using the equation $\mu_0H_{c1}(T)=\mu_0H_{c1}(0)(1-(T/T_c)^2)$. The estimated $\mu_0H_{c1}(0)$ at zero temperature of $CuIr_{1.925}Al_{0.075}Te_4$ sample is 0.060 T, which is larger than that of undoped parent sample (see in **Table 2**).

To further convince that SC is essential feature of the highest $T_c$ compound $CuIr_{1.925}Al_{0.075}Te_4$, we also perform the temperature-dependent specific heat measurement. **Fig. 5** illustrates the detailed characterization of the superconducting transition in the highest $T_c$ composition $CuIr_{1.925}Al_{0.075}Te_4$ through specific heat measurements under absence of magnetic field. The data $C_p/T$ *vs.* $T^2$ can be fitted by the equation $C_p = \gamma T + \beta T^3$ above the $T_c$ to acquire the value of $\beta$ and $\gamma$ is 2.20 mJ mol$^{-1}$ K$^{-4}$ and 12.12 mJ mol$^{-1}$ K$^{-2}$ for $CuIr_{1.925}Al_{0.075}Te_4$, respectively, where $\gamma T$ represents the sum of electron contributions ($C_{el.}$) to the specific heat and $\beta T^3$ is the sum of phonon contributions ($C_{ph.}$). **Fig. 5b** shows the sum of electron contribution to the specific heat at the temperature near the $T_c$ under 0 T, where $C_{el.}$ is easily derived from deducting the phonon part: $C_{el.} = C_p - \beta T^3$. Apparently, a sharp specific heat jump occurs in our representative $CuIr_{1.925}Al_{0.075}Te_4$, characteristic of bulk SC. The $T_c$ further can be determined to be 2.70 K using the common equal-area entropy construction method, which agrees well with those observed in magnetization and resistivity tests. Based on the $T_c$ and $\gamma$ values, we can determine $\frac{\Delta C_{el.}}{\gamma T_c} = 1.53$, which is slightly larger than the value of 1.43 forecasted by the BCS theory, revealing its superconducting nature. The Debye temperature obtained using the equation $\Theta_D = (12\pi^4 nR/5\beta)^{1/3}$ is 183.5(1) K, where $R$ represents the gas constant, $n$ expresses the number of atoms per formula unit. The resultant electron-phonon coupling constant $\lambda_{ep}$ value further estimated by introducing the $\Theta_D$ number into the inverted McMillan equation: $\lambda_{ep} = \dfrac{1.04 + \mu^* \ln\left(\frac{\Theta_D}{1.45 T_c}\right)}{(1-0.62\mu^*)\ln\left(\frac{\Theta_D}{1.45 T_c}\right) - 1.04}$ [49] for $CuIr_{1.925}Al_{0.075}Te_4$ is 0.66. Following the formula $N(E_F) = \dfrac{3}{\pi^2 k_B^2 (1+\lambda_{ep})}\gamma$, the $\lambda_{ep}$ and $\gamma$ values can give rise to the electron density of states (DOSs) around the Fermi level ($N(E_F)$). The obtained $N(E_F) = 3.13$ states/eV f.u. for $CuIr_{1.925}Al_{0.075}Te_4$, greater than that of pristine $CuIr_2Te_4$ (($N(E_F) = 3.10$ states/eV f.u). The enhancement of $T_c$ may be accompanied by the increment of the DOS states at the Fermi energy in $CuIr_{2-x}Al_xTe_4$ systems. As show in **Fig. 5b**, the specific heat data can be basically fit with the equation $C_{el.} = A \exp(-\Delta/(k_B T))$. Due to the limitation of experimental conditions, the specific heat measurement did not measure the lower temperature region. Nevertheless, the $R^2$ value of 0.9998 shows that the data and the calculated curve are reasonably consistent.

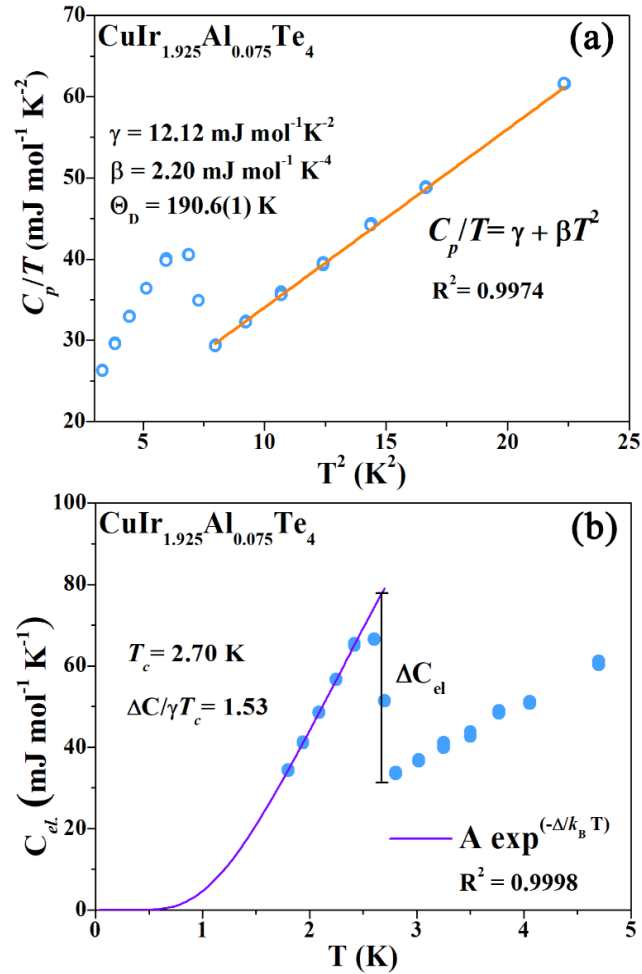

**Fig. 5** (a) Specific heat data plotted as a function of $T^2$ for $CuIr_{1.925}Al_{0.075}Te_4$. (b) the electronic part of the specific heat in $CuIr_{1.925}Al_{0.075}Te_4$.

Finally, to further understand the effect of Al dopant on the CDW and SC of $CuIr_2Te_4$, the electronic phase diagrams plotted $T_c$ versus Al-content $x$ have been established (see **Fig. 6**). All the $T_c$ values were obtained from the resistivity and magnetization tests. From **Fig. 6**, we can find that the CDW order is immediately suppressed while $T_c$ increases with increasing Al content up to $x = 0.075$ and rises towards the highest value 2.75 K at $x = 0.075$. From this, we find that despite subtle Al is instead of Ir, it has a strong impact on the SC and CDW. Besides, in contrast to our previously reported system Ru/Ti-doped $CuIr_2Te_4$, the similarity is the destabilization of CDW upon small amount doping concentration no matter Ru, Ti, or Al as dopants and formation of dome-shape like superconducting phase diagrams. [32,33] Despite overall similarities, there are significant differences between the $CuIr_{2-x}Ru_xTe_4$, $CuIr_{2-x}Ti_xTe_4$, and $CuIr_{2-x}Al_xTe_4$ systems. Substitution of Ir by Ru or Ti in $CuIr_2Te_4$ corresponds to a "hole" ($p$-type) doping of the $IrTe_2$ layers, while partial doping Al into Ir site in $CuIr_2Te_4$ is an electron ($n$-type) doping. As for the high-$T_c$ cuprate superconductors, where the competition between anti-ferromagnetism and SC develops as a function of chemical doping, the evolving balance between competing electronic orders in CDW/SC systems is one of their most fundamentally tempting properties. In our case, the bands in the neighborhood of the Fermi energy $E_F$ of the pristine sample $CuIr_2Te_4$ mostly come from Ir $d$ and Te $p$ orbitals and locate at a flat plateau, in which Al doping acts as a chemical pressure, closing the gap on the Fermi surface that usually leads to rapid suppression of CDW. On the other hand, the increase of band filling of the Fermi surface under chemical pressure could be the mechanism of the promotion of SC. But further study and evidence need to be collected to find out the competition between CDW and SC.

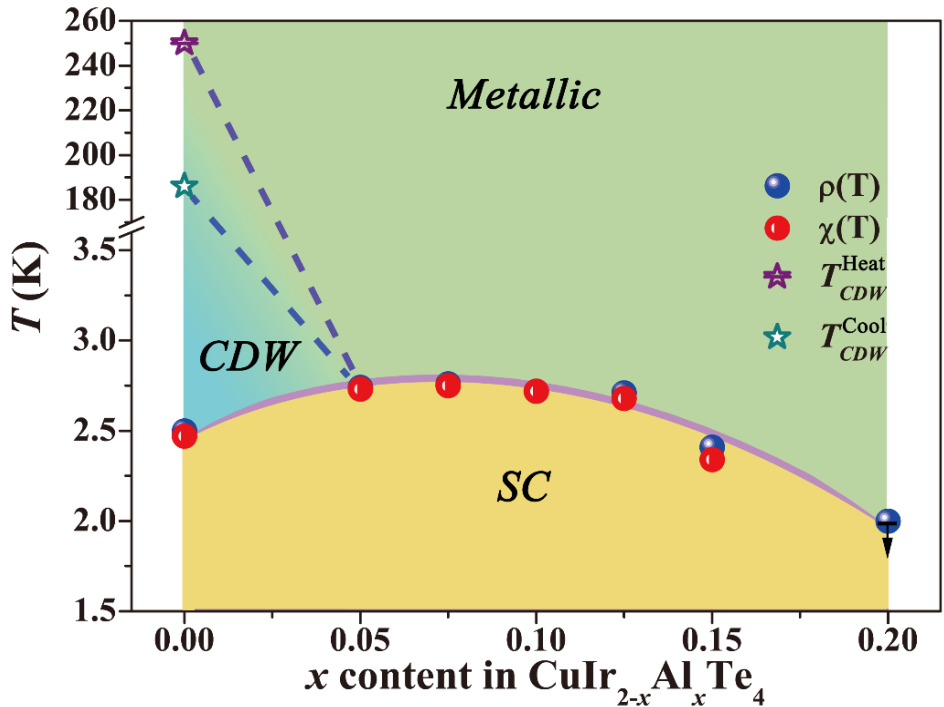

**Fig. 6** The electronic phase diagram of polycrystalline $CuIr_{2-x}Al_xTe_4$ ($0 \leq x \leq 0.2$) series.

4. **Conclusion and perspectives**

In conclusion, we have successfully synthesized a series of polycrystalline CuIr$_{2-x}$Al$_x$Te$_4$ ($0 \leq x \leq 0.2$) samples via a solid-state method and systemically studied the effect of Al doping on the structure and electronic properties of CuIr$_2$Te$_4$. The $\frac{\Delta C_{el.}}{\gamma T_c} = 1.53$ for the highest $T_c$ sample CuIr$_{1.925}$Al$_{0.075}$Te$_4$ is slightly larger than 1.43 (BCS value), proving its bulk superconducting nature. We recognize that the CDW order is suppressed immediately while $T_c$ increases as Al doping amount $x$ rises and achieves a maximum $T_c$ = 2.75 K with Al doping content of 0.075. Our systematic study of CuIr$_{2-x}$Al$_x$Te$_4$ ($0 \leq x \leq 0.2$) not only extends the family of the TMD superconductors but also provides a platform for further research on the relationship between the CDW and SC.